\documentclass[aps,prl,amssymb,showpacs,twocolumn]{revtex4}
\usepackage{graphicx}

\begin{document}

\title{Photon Mixing in Domain Walls and the Cosmic Coincidence Problem}

\author{Jarah Evslin}
\email{jevslin@ulb.ac.be}
\affiliation{Universite Libre de Bruxelles, Campus de la Plaine, C.P. 231, Boulevard du Triomphe, B-1050 Bruxelles, Belgium}
\author{Malcolm Fairbairn}
\email{malc@physto.se}
\affiliation{Cosmology, Particle astrophysics and String theory, Department of Physics, Stockholm University, AlbaNova University Centre, SE-106 91, Stockholm, Sweden}


\newcommand{\lP}{\ell_{\mathrm P}}

\newcommand{\md}{{\mathrm{d}}}
\newcommand{\Kern}{\mathop{\mathrm{ker}}}
\newcommand{\tr}{\mathop{\mathrm{tr}}}
\newcommand{\sgn}{\mathop{\mathrm{sgn}}}

\newcommand*{\R}{{\mathbb R}}
\newcommand*{\N}{{\mathbb N}}
\newcommand*{\Z}{{\mathbb Z}}
\newcommand*{\Q}{{\mathbb Q}}
\newcommand*{\C}{{\mathbb C}}

\begin{abstract}
A model is presented where there exists another U(1) gauge group which is extremely weakly coupled to that of QED except inside the core of domain walls.  It is possible to choose parameters such that standard model photons crossing such a wall are mixed maximally with the 'para-photons' of the other U(1).  We use this model to explain the apparent low luminosity of high redshift supernovae.  A possible mechanism is outlined where the domain walls correspond to changes in the relative orientation of branes in a compact space.  The model is rather contrived but has the advantage that for a wide range of values it can solve the cosmic coincidence problem - an observer at any redshift would come to the conclusion that their universe had recently started to accelerate.
\end{abstract}

\maketitle

Up until relatively recently the cosmological constant problem was simply a question of why the vacuum energy was approximately zero \cite{weinberg} but over the past decade the problem has developed a disturbing new characteristic - reconstructions of the Hubble diagram using observations of type 1a supernovae suggest that the universe has recently started to accelerate at a redshift of approximately unity (see e.g. \cite{knop}).  This suggests that not only is the cosmological constant extremely small, it is finely tuned to be extremely close to the energy density of matter in the universe today.

There are various candidate solutions to this problem in the literature, the most famous being tracking quintessence models \cite{quintessence} but these require rather finely tuned potentials with extremely small masses of the order of the inverse radius of the universe.  Anthropic arguments are rather powerful \cite{weinberg2} but it is interesting, even if just as an exercise, to see if there are any alternatives before we resort to such arguments which would after all see us turning our back on the scientific method.

One way of turning the problem around is to interpret the dimness of high redshift supernovae as being due to a dimming of the photons coming from those objects. Dimming due to dust is strongly constrained since there is no evidence for a redshift dependence of the colour of supernovae and dust tends to distort spectra \cite{dust}.  An attractive alternative is the dimming of photons due to their mixing with a light pseudo-scalar axion in the intergalactic magnetic field \cite{csaki}.  However, the following arguments might be raised against such models:  
In order to get the right ammount of photon dimming, the models require a particular value for both the axion coupling and the strength of the intergalactic magnetic field $\sim 10^{-9}$ G which some studies suggest is an overestimate \cite{tkachev}.  Furthermore, the presence of electrons in the intergalactic medium makes the dimming dangerously frequency dependent \cite{frequency}.  Finally, photon/axion mixing does not solve the coincidence problem - there is no particular reason why the dimming should occur on a length scale equal to the size of the universe {\it today}.

The aim of this letter is to present a new scenario which, while obviously borrowing heavily from the axion-photon model, has some important distinctions which may overcome the objections listed above.  The main requirements for such a mechanism are that the dimming of photons takes place on a length scale tied to the size of the horizon and that dimming of light from more distant objects must saturate since we see photons from very high redshifts.   

It is possible that there exist para-photons associated with other unbroken U(1) gauge symmetries in the particle spectrum, provided they are very weakly coupled to our photons \cite{okun,paraphoton}.  The presence of another U(1) would lead to a kinetic mixing term between our photons and the para-photons of the form
\begin{equation}
\mathcal{L}=-\frac{1}{4g^2}F^{\mu\nu}F_{\mu\nu}-\frac{1}{4g'^2}G^{\mu\nu}G_{\mu\nu}-\frac{\chi}{g'g} F^{\mu\nu}G_{\mu\nu}
\label{lagr}
\end{equation}
where $F_{\mu\nu}=\partial_\mu A_\nu-\partial_\nu A_\mu$ is the normal field strength of electromagnetism and $G_{\mu\nu}=\partial_\mu A'_\nu-\partial_\nu A'_\mu$ is the field strength associated with the para-photon's U(1). 

The photons produced by an accelerating electron will be linear superpositions of our photon and the para-photon \cite{okun,foot} so there will be a physical photon $A_1$ which is a superposition of $A$ and $A'$ and a sterile photon $A_2$ which is orthogonal to $A_1$.  The two states are given by
\begin{eqnarray}
A_1=\frac{A+2\chi g'g A}{\sqrt{1+(2\chi g'g)^2}}\nonumber\\
A_2=\frac{A'-2\chi g'g A}{\sqrt{1+(2\chi g'g)^2}}
\end{eqnarray}

Mixing between the two states may occur although there are some important differences between this situation and the the axion/photon case.  In particular, if the photons remain massless, mixing will only occur if coherent forward scattering off electrons or para-electrons changes the refractive index of each photon species differently.  In medium interactions will then be able to change the amount of $A$ and $A'$ in the propagating wave. The mixing length $l_{osc}$ will be given by (see e.g. \cite{wolfenstein})
\begin{equation}
l_{osc}\sim \left|\frac{m_e^2g^4}{n_e}-\frac{m_{e'}^2 g'^4}{n_{e'}}\right|
\end{equation}
and the mixing angle by
\begin{equation}
\sin^2 (2 \theta) \sim \chi g'g.
\end{equation}
Since we keep the para-photons and the photons massless, there is no frequency dependence in the mixing which allows us to evade some of the constraints on photon/axion mixing which come from the non-observation of colour change in supernova spectra \cite{goob}.  Also, since each of the two polarisations of the photon has its own para-photon degree of freedom to mix with, there will be no dilution of polarisation over distance.

Such an interaction between our photons and another U(1) field is tightly constrained by nucleosynthesis and the absence of millicharged particles \cite{raffelt} so such a mixing must be very small on earth and throughout the majority of space.  However, it is not possible to say if this is true everywhere in space, in particular we will suggest that domain walls exist within which the mixing angle becomes much larger, close to unity.

As the universe cools through phase transitions, topological objects can be formed which interpolate between causally disconnected regions which have made different choices as to their low energy vacuum configurations.  Depending on the topology of the vacuum networks of strings or domain walls may form. Work by Kibble predicts that there will be approximately one such defect per horizon size \cite{kibble} and studies have shown that for strings and domain walls the correlation length of defect networks grows approximately linearly with time (see e.g.\cite{scaling}), as does the horizon size in the post inflationary universe, so at any given time there will be of order one defect per horizon size.

Consider the situation now that there is a scaling domain wall network throughout the universe and the field responsible for domain wall configuration is the same one which sets the coupling between the standard model photons and the other U(1) gauge symmetry. The domain walls are such that there is large coupling between our photons and the para-photons inside the core of the domain wall and there is a non-zero electron/para-electron density.  We assume that the probability of a physical $A_1$ photon becoming a sterile $A_2$ photon as it crosses a single wall is 0.5, and vice versa.  

The number of domain walls that will be passed by a photon emitted at time $t_{e}$ (redshift $z_e$) and arriving on earth today at $t_0$ can be approximated by the expression
\begin{equation}
N(z)=a_0\int_{t_e}^{t_0}\frac{dt}{a(t)\zeta(t)}=\frac{2z_e}{3k}
\end{equation}
where we assume the correlation length $\zeta(t)=kt$ where $k$ is some constant of order 1.  The luminosity distance will be changed in the following way \cite{grojean}
\begin{equation}
d_L(z)=(1+z)r(z)a_0/P_{\gamma\rightarrow\gamma}^{1/2}
\end{equation}
 where $r(z)$ is the usual expression for the co-moving distance to an object at a redshift of $z$ \cite{grojean} and the probability of the photon remaining a photon and not mixing into a para-photon $P_{\gamma\rightarrow\gamma}$ is given by
\begin{equation}
P_{\gamma\rightarrow\gamma}=0.5\left(1+e^{-N(z)}\right)
\end{equation}

The length scale of the dimming is therefore always approximately equal to the size of the universe, as required in order to explain the coincidence problem.

\begin{figure}[h]
\includegraphics[scale=0.65]{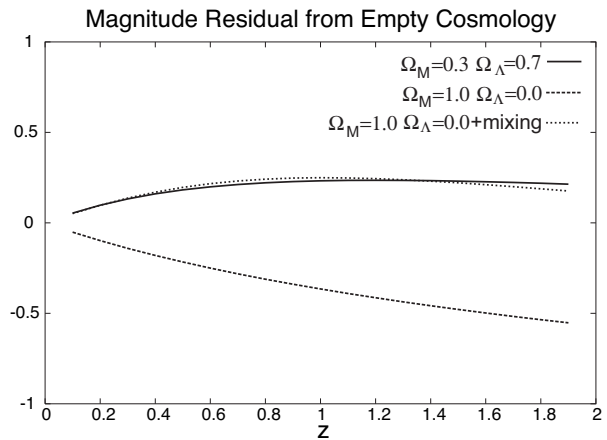}
\caption{\label{sn}The effect of mixing in domain walls on the Hubble Diagram.  Here plotted are the residual magnitudes relative to an empty Milne Universe for the concurrence model $\Omega_M=0.3,\Omega_\Lambda=0.7$, the disfavoured flat, matter dominated universe $\Omega_M=1.0,\Omega_\Lambda=0.0$ and the same flat universe with photon mixing in domain walls. In this plot, we assume the domain wall correlation length $\zeta(t)=kt$ with $k=1/3$ and a probability of 0.5 for mixing in each wall.}
\end{figure}

Photons coming from the CMB would also pass through domain walls of course, and fluctuations in the number of photons arriving will occur.  This seems dangerous since we know that the fluctuations in the temperature anisotropy are extremely small, corresponding to variations in the number of photons arriving of $\Delta n/n < 10^{-5}$.  However, we know that photons coming from the CMB will cross many ($\sim 1000$) domain walls before they reach the earth and even though each crossing will correspond to some photons being converted into para-photons, after some time as many para-photons should become photons as vice versa.

\begin{figure}[h]
\includegraphics[scale=0.65]{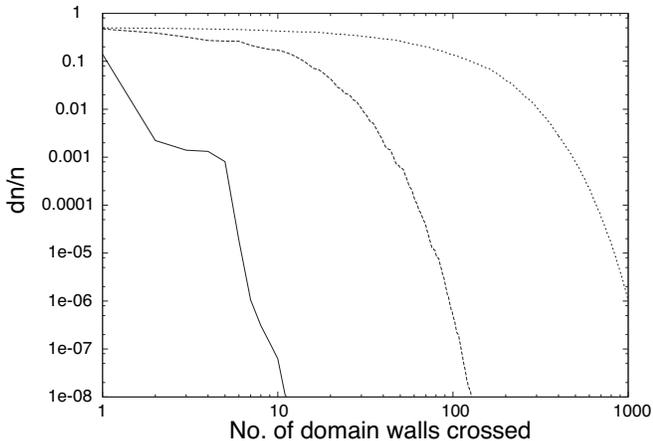}
\caption{\label{cmb}Plot showing the effect on the CMB of domain wall crossing. The vertical scale is the random variation in the total number of photons arriving induced by the domain wall crossings (see text). The three plots correspond to the mixing probabilities 0.5, 0.05 and 0.005 in a single wall.}
\end{figure}

Simple numerical estimates show this is true, figure \ref{cmb} shows the induced effect on the perturbations in the CMB due to mixing into para-photons as they cross domain walls.  The most rapidly dropping plot in that diagram corresponds to a mixing probability of 0.5 per domain wall and shows that the induced perturbation drops below the magnitude of the initial real perturbation after around 5 crossings.

Domain walls are sheets of vacuum energy which stretch across the universe, and consequently their gravitational effect can be very strong, in particular creating larger anisotropies in the CMB than those observed \cite{zeldovich}.  This constraint can be written
\begin{equation}
\sigma H \sim V_0 w H \ll 10^{-5}\rho_{crit}
\end{equation}
where $\sigma$ is the mass per unit area of the domain wall, $H\sim 10^{33} eV^{-1}$ is the Hubble constant, the inverse of which sets the size of the domain wall, $V_0$ is the energy density inside the core of the domain wall and $w$ is the width of the wall. $\rho_{crit}\sim (10^{-3}$ eV$)^{4}$ is the critical density of the universe, which leaves us with the inequality
\begin{equation}
\sigma\sim V_0 w\ll 10^{16} eV^3
\end{equation}

In order for the probability of the photon disappearing to be around 0.5 in each domain wall crossing, we need to arrange the thickness of the domain wall $w$ to be at least as large as the mixing length inside the wall $l_{osc}$.  The density of electrons in intergalactic space is rather small, so it is easier to use scattering from para-electrons to create mixing since their mass and hence number density and cross section is much less constrained.  The basic relationship which needs to be satisfied is
\begin{equation}
w > \frac{m_{e'}^2 g'^4}{n_{e'}}
\end{equation}
where the density of para-electrons is constrained to be $n_{e'}m_{e'}\ll \rho_{crit}$, unless one was to suggest that they are a significant fraction of dark matter.

We have not yet presented a possible candidate for domain walls inside which the coupling between the two U(1) groups becomes very large, whereas outside the coupling is small.  One possibility is for the U(1) groups to be located on two branes which fill our three decompactified dimensions but are separated by a finite distance in a compact higher dimensional space due to their mutual repulsion.  Such an interbrane potential can be motivated in the context of the potential between two D-branes in string theory, see e.g. \cite{dvali} where it would take the form
\begin{equation}
V(r)=\tau\left(\alpha+\sum_i \beta_i\frac{e^{-m_i r}}{(r/l_s)^{N-2}}\right)
\end{equation}
where $r$ is the distance between the branes, $N$ is the number of codimensions in which the interbrane fields can propagate, $\tau$ is the tension of the brane, $l_s$ is the string length and $\alpha$ is some parameter which sets the cosmological constant to zero in vacuum by some other mechanism. The index $i$ represents the sum over fields in the bulk of mass $m_i$ and $\beta_i$ are model Dependant constants.  At least some fields in the bulk will have obtained a non-zero mass due to supersymmetry breaking which must occur on our brane.  We neglect the possible effect of any massive strings stretching between the branes, and also any tachyonic instability, since we assume that the distance between the branes remains much larger than $l_s$.

\begin{figure}[h]
\includegraphics[scale=0.35]{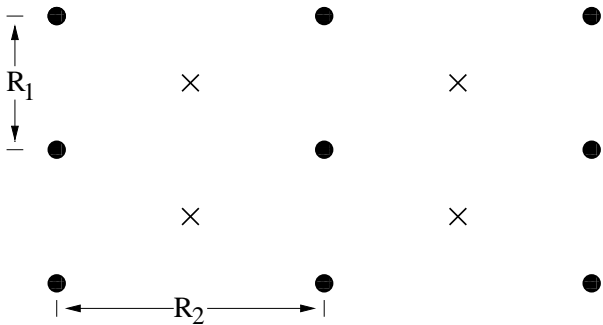}
\caption{\label{compact} Schematic diagram of the two mutually repulsive branes in two large extra dimensions where 4 copies of the compact space are shown.  The brane containing the standard model U(1) is labeled by a cross and the brane containing the para-photon U(1) is denoted by a dot.}
\end{figure}

We assume that compact space is a product of tori with radii fixed by mechanisms other than a minimum in the interbrane potential.  If there is one or more dimension of radius $R_1$ and another with a much larger radius $R_2$ then the two branes will arrange themselves to minimise the free energy and hence get as far apart as possible given the periodicity of the space, as shown schematically in figure \ref{compact}.  
If there are only two co-dimensions into which the interbrane fields can propagate perpendicular to the brane world volume then zero mass fields would form a flux tube at distances $r>R_1$ that would not diminish with distance so if the space is compact, there would be no potential unless the fields get a mass.

If at some earlier time in the universe the branes had kinetic energy associated with temperature which was larger than the energy associated with the repulsive potential then domain walls would be created as the universe cooled.  The domain wall configuration would correspond to a brane moving between one lattice sight and the adjacent lattice sight in the $R_2$ direction during which the standard model U(1) (cross) brane moves close to the para-photon U(1) (dot) brane as it passes it.  Kinetic mixing between the branes occurs via the exchange of bulk modes \cite{kmix} so if the modes responsible for the mixing are massive, then we can get strong mixing in the core of the domain walls and very suppressed mixing outside the domain wall, provided that $R_2\gg m^{-1}\ge R_1$.

At the first approximation, the mass scale of the fields in the bulk $m$ will set the inverse width of the domain wall $w\sim m^{-1}$ and the energy density inside the core of the domain wall will be set by the tension of the brane $\tau$.  Phenomenologically the only constraint on the tension of the brane would be presumably be the same as for light scalars minimally coupled to gravity, in other words the mass would have to be larger than $10^{-3}$ eV in order to avoid discrepancies with tests of Newton's law at short distances.  However theoretically such a small brane tension may be a problem in string theory, since it is difficult to get brane tensions with energy densities lower than the string scale which is constrained by collider experiments to be at least greater than a few hundred GeV.  There are branes in string theory with low tensions, for instance branes wrapped around compact loops which shrink to zero size \cite{strominger} but this question will have to be addressed in more detail later.  

In conclusion, a model has been presented where our standard model photons are strongly mixed with para-photons each time they cross a domain wall.  Since there is approximately one domain wall per horizon size, supernovae are dimmed on a length scale associated with the horizon size at any redshift.  This solves the cosmological coincidence problem.

The domain walls required are very unusual in that they require a very low mass per unit area, and photon/para-photon mixing has to be very strong withing the walls and very much suppressed outside.  However, unlike other models, non of those parameters have any relation to the present energy density or size of the universe.

Of course we are well aware of the discrepancy between the value for $\Omega_M\sim 0.3$ obtained from galaxy clustering and the $\Omega_{total}\sim 1$ obtained from the position of the first peak in the CMB - we have not attempted to address that problem here.  We simply wish to point out that the supernova data may have some explanation other than an extremely finely tuned vacuum energy via some mechanism in the spirit of that outlined above.

{\bf Acknowledgments} It is a pleasure to thank the following people for their patience and comments -  Pisin Chen, Martin Ericsson, Ariel Goobar, Mark Hindmarsh, Sasha Ignatiev, John Mather, Edvard M\"ortsell, Timur Rashba, Georg Raffelt and Jon Terning.

\end{document}